\documentclass[twocolumn,preprintnumbers,amsmath,amssymb]{revtex4}
\usepackage{subfigure}
\usepackage{graphicx}
\usepackage{dcolumn}
\usepackage{bm}

\begin{document}

\title{Competing Alignments of Nematic Liquid Crystals on Square Patterned Substrates}
\author{C. Anquetil-Deck}
\email[Present address: Karlsruhe Institute of Technology, Institute for Meteorology and Climate Research, Atmospheric Aerosol Research Department (IMK-
AAF), Hermann-von-Helmholtz-Platz 1, D - 76344 Eggenstein-Leopoldshafen
Germany.]{}
\author{D.J. Cleaver}
\email{d.j.cleaver@shu.ac.uk}

\affiliation{ Materials and Engineering Research Institute, Sheffield Hallam
University, City Campus, Howard Street, Sheffield, S1 1WB,UK }

\author{T.J. Atherton}

\affiliation{ Department of Physics and Astronomy, Tufts University, 4 Colby Street, Medford, Massachusetts, USA 02155 }

\date{\today}

\begin{abstract}
A theoretical analysis is presented of a nematic liquid crystal confined
between substrates patterned with squares that promote vertical and
planar alignment. Two approaches are used to elucidate the behavior
across a wide range of length scales: Monte Carlo simulation of hard
particles and Frank-Oseen continuum theory. Both approaches predict
bistable degenerate azimuthal alignment in the bulk along the edges
of the squares; the continuum calculation additionally reveals the
possibility of an anchoring transition to diagonal alignment if the
polar anchoring energy associated with the pattern is sufficiently
weak. Unlike the striped systems previously analyzed, the Monte Carlo
simulations suggest that there is no {}``bridging'' transition for
sufficiently thin cells. The extent to which these geometrically patterned
systems resemble topographically patterned substrates, such as square
wells, is also discussed.
\end{abstract}

\maketitle

\section{Introduction}

The imposition of a liquid crystal's (LC's) bulk director orientation through that LC's interaction with a confining substrate is termed anchoring~\cite{RefWorks:2}. In the absence of defects and applied fields, substrate anchoring is the main determinant of the director profile in a sandwich geometry LC cell; the director profile in such a cell is set through 0 of the orientational elastic energy, subject to each wall's polar and azimuthal anchoring constraints. At a continuum level, this elastic energy is most commonly expressed through square director gradient terms corresponding to the independent splay, twist and bend modes of orientational deformation, weighted by the elastic constants $K_{1}$, $K_{2}$ and $K_{3}$, respectively. When considered at a finer length-scale, conversely, both the anchoring strengths and the bulk elastic constants are emergent from the microscopic interplay of the orientational and positional degrees of freedom of the liquid crystalline molecules and the confining surfaces. 

Traditional routes to establishing desired anchoring 0, and, thus, director profiles of use for display devices, include substrate rubbing and various photo-alignment approaches (light-induced cis-trans isomerisation, photodegradation ...). Whilst significant empirical knowledge has been developed in relation to each of these approaches, the molecular mechanisms by which they achieve desired 0 are poorly understood. Introducing inhomogeneity into substrate conditions has, for some time, been 0 as an alternative route to both controlling conventional anchoring and, increasingly, introducing new phenomena. A range of such substrates have been developed and examined. These cover patterning length-scales ranging from $10^{-7}$m upwards, couple to the LC either sterically, chemically or dielectrically (or by a combination of same) and have been achieved as both 1-dimensional (stripes, ridges, ...) and 2-dimensional (circles, squares, triangles, posts, ... ) patternings.

One of the important phenomena that can be achieved through substrate patterning is bistability, that is 0 of two distinct anchoring arrangements with (in the absence of an applied field) mutually inaccessible free energy minima. Pattern-0 0 has now been established for the blazed grating structure of the Zenithally Bistable Device~\cite{RefWorks:3,Jones2003}, the two-dimensional array of the Post Aligned Bistable Nematic~\cite{RefWorks:4} device and, more recently, a steric square-well arrangement~\cite{RefWorks:5,RefWorks:6}. In each of these, the bistability pertains between one state with a continuous director arrangement and a second containing orientational defects that are pinned in some way by the substrate inhomogeneity. This suggests that the key length-scale for achieving bistability here is the size and periodicity of the patterning, a conjecture which is supported by the success of mesoscopic modelling approaches in both accessing the bistable states and relating, semi-quantitatively, switching fields to geometrical parameters \cite{RefWorks:7,PhysRevE.82.041703}.

In addition to these sterically patterned systems, chemical patterning has now also been developed as an approach for imposing substrate inhomogeneity on LC systems. The notion of imposing combinations of azimuthal and polar anchorings on LCs via chemically nanopatterned substrates was the subject of early experimental work~\cite{RefWorks:8,RefWorks:9,RefWorks:10}.
Subsequently, Lee and Clark performed a more systematic study of the alignment properties of nematic LCs on
surfaces comprising both homeotropic and planar alignment areas~\cite{RefWorks:16}. For stripe patterns, they found that the polar orientation depends on the relative areas of the homeotropic and planar regions, but that the azimuthal anchoring always runs along the direction of the stripes. In 2005, Scharf $et$ $al.$~\cite{RefWorks:17,RefWorks:18} undertook further investigations of systems with competing alignment regions.
Further innovations by the groups of Abbott~\cite{RefWorks:19} and Evans~\cite{RefWorks:21} 0 on the development of patterns of combinations of alkanethiols deposited as self assembled monolayers (SAMs) on gold. Using microcontact printing, these systems proved capable of achieving highly reproducible surface features with periodicities of 10s of microns. Square, circular and striped patterns written on these length-scales were, thus, observed using optical microscopy in crossed 2 set-ups. An alternative approach, employing selective ultra-violet
irradiation of SAMs, achieved LC-aligning stripe patterns on the sub-micron scale~\cite{RefWorks:22}.

In Ref~\cite{RefWorks:23}, two of the current authors contributed the simulation aspects of a joint experimental / simulation study of LC alignment at a single patterned substrate. In this, it was shown that a range of patterned SAMs can be used to control LC alignment states and domains. For stripe patterns, the LC was found to align parallel to the
stripe boundaries for both nanoscale simulation features and micron-scale experimental systems.
Indeed, despite the significantly different length-scales involved, the qualitative 0 seen
in simulations of generic molecular models confined using a striped substrate proved entirely consistent with the experimental observations. Specifically, on undergoing isotropic to nematic ordering, all systems proved to be dominated by the homeotropic-aligning substrate regions {\em at} the ordering transition, the influence of the planar-aligning regions only becoming apparent well into the nematic phase.

In Ref~\cite{RefWorks:24}, we extended our molecular simulation work to consider the Ã0 of a thin nematic film confined between two identical nano-patterned substrates. Using patterns involving alternating stripes of homeotropic-0 and homogeneous-0 substrate, we showed that the polar anchoring angle can be varied continuously from planar to homeotropic by appropriate tuning of the relative stripe widths and the film thickness. For thin films with equal stripe widths, we also observed orientational bridging, the surface patterning being written in domains which traversed the nematic film. This dual-bridging-domain arrangement broke down with increase in film thickness, however, being replaced by a single tilted monodomain. Strong azimuthal anchoring in the plane of the stripe boundaries was observed for all systems.

Stripe-geometry systems have also been analyzed by the third of the current authors using continuum theory~\cite{RefWorks:11,RefWorks:13,RefWorks:12}. This larger length-scale work has shown that the basis for azimuthal alignment by striped substrates is associated with differences in the Frank elastic constants. Azimuthal anchoring parallel to the stripes corresponds to the LC adopting a configuration comprising twist, splay and bend deformations; in the other limiting case, bulk alignment perpendicular to the stripes, only splay and bend deformations are required. Experimentally, $K_{2}$ is significantly lower than $K_{1}$ and $K_{3}$ for most nematics, so that parallel anchoring is stable. Monte Carlo estimates of the elastic constants for calamitic particle-based LC simulation models yield similar elastic constant ratios~\cite{RefWorks:25}, so this phenomenological agreement between the predictions of particle-based and continuum approaches is to be expected.

In this paper, we extend our respective works on stripe-patterned systems by studying the effect of substrates with square patternings on a confined LC film. Experimental studies of such systems include the checkerboard patternings achieved by Bramble~\cite{RefWorks:23} and, more recently, Yi~\cite{RefWorks:26} and the bistable square-well systems mentioned above~\cite{RefWorks:5,RefWorks:6}. In respect to the latter, we note that both Q-tensor~\cite{Cornford2011} and Landau-De Gennes~\cite{RefWorks:28} modelling approaches have been used to examine the stable configurations for such systems. From this, diagonally-anchored and edge-anchored states have been identified the former comprising surface region defects.

Here we use both molecular- and continuum-level modelling approaches to investigate the 0 of LC films confined between square-patterned substrates. In section II we present our molecular-level model system and describe the simulation methodology employed. Section III contains the corresponding simulation results. Following this, in Section IV we present a continuum-level analysis of anchoring control in systems with square-patterned substrates. Finally, in Section V, we compare and combine the findings from these investigations to draw more general conclusions.

\section{Molecular model and simulation details}
We have performed a series of Monte Carlo (MC) simulations of
rod-shaped particles confined in slab geometry between two planar
walls. Inter-particle interactions have been modelled through the
Hard Gaussian Overlap (HGO) potential \cite{RefWorks:29}. Here, the
dependence of the interaction potential $\nu^{HGO}$ on ${\hat{\bf
u}}_{i}$ and ${\hat{\bf u}}_{j}$, the orientations of particles
$i$ and $j$, and ${\hat{\bf r}}_{ij}$, the inter-particle unit
vector is
\begin{equation}
\nu^{HGO} = \left\{ \begin{array}{ll}
0 & \textrm{if $r_{ij}\ge\sigma({\hat{\bf r}}_{ij},{\hat{\bf u}}_{i},{\hat{\bf u}}_{j})$}\\
\infty & \textrm{if $r_{ij}<\sigma({\hat{\bf r}}_{ij},{\hat{\bf u}}_{i},{\hat{\bf u}}_{j})$}\\
\label{UHGO}
\end{array} \right.
\end{equation}
where $\sigma({\hat{\bf r}}_{ij},{\hat{\bf u}}_{i},{\hat{\bf
u}}_{j})$, the contact distance, is given by
\begin{widetext}
\begin{eqnarray}
\sigma({\hat{\bf r}}_{ij},{\hat{\bf u}}_{i},{\hat{\bf
u}}_{j})=\sigma_{0}\left[1-\frac{\chi}{2}\left[\frac{\left({\hat{\bf
r}}_{ij}.{\hat{\bf u}}_{i}+{\hat{\bf r}}_{ij}.{\hat{\bf
u}}_{j}\right)^{2}}{1+\chi ({\hat{\bf u}}_{i}.{\hat{\bf
u}}_{j})}+\frac{\left({\hat{\bf r}}_{ij}.{\hat{\bf
u}}_{i}-{\hat{\bf r}}_{ij}.{\hat{\bf u}}_{j}\right)^{2}}{1-\chi
({\hat{\bf u}}_{i}.{\hat{\bf u}}_{j})}\right]\right]^{-1/2} .
\label{sigma}
\end{eqnarray}
\end{widetext}
The parameter $\chi$ is set by the particle length to breadth
ratio $\kappa$= $\sigma_{end}/\sigma_{side}$ via
\begin{equation}
\chi=\frac{\kappa ^{2}-1}{\kappa ^{2}+1} .
\end{equation}

\begin{figure}[!h]
\begin{center}
\includegraphics[scale=0.3]{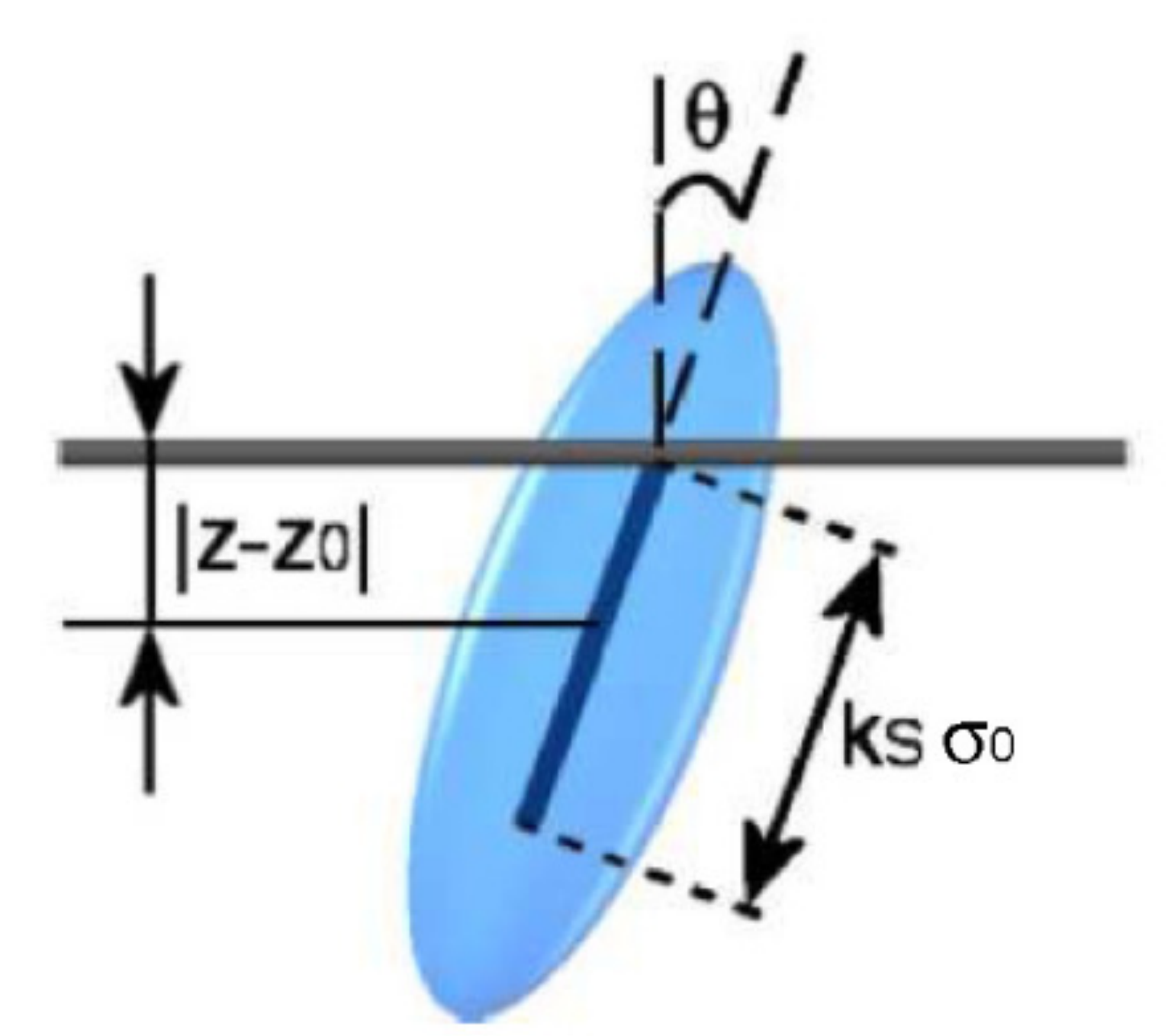}
\caption{(Color online)Schematic representation of the geometry used for the
hard needle-wall (HNW) particle-substrate interaction
\cite{RefWorks:30}.} \label{hnw}
\end{center}
\end{figure}

Particle-substrate interactions have been modelled using the hard
needle-wall potential (HNW) \cite{RefWorks:30}. In this, the
particles do not interact directly with the surfaces. Rather the
surface interaction is achieved by considering a hard axial needle
of length $\sigma_{0}k_{s}$ placed at the center of each particle
(see Figure ~\ref{hnw}). This gives an interaction
\begin{equation}
\nu^{HNW} = \left\{ \begin{array}{ll}
0 & \textrm{if $|z_{i}-z_{0}|\ge\sigma_{w}({\hat{\bf u}}_{i})$}\\
\infty & \textrm{if $|z_{i}-z_{0}|<\sigma_{w}({\hat{\bf u}}_{i})$}\\
\label{UHNW}
\end{array} \right.
\end{equation}
where $z_{0}$ represents the location of a substrate and
\begin{equation}
\sigma_{w}({\hat{\bf
u}}_{i})=\frac{1}{2}\sigma_{0}k_{s}\cos(\theta_{i})
\end{equation}
Here, $k_{s}$ is the dimensionless needle length and
$\theta_{i}=\arccos(u_{i,z})$ is the angle between the substrate
normal and the particle's orientation vector. For small $k_{s}$, the
homeotropic arrangement has been shown to be stable, whereas
planar anchoring is 0 for long $k_{s}$
\cite{RefWorks:30}. Furthermore, despite its simplicity, the
HNW potential has been found to exhibit qualitatively identical
0 to that obtained using more complex particle-substrate
potentials \cite{RefWorks:14}. Here, by imposing variation in
$k_{s}$ across the two boundary walls, we investigate the effects of
molecular-scale substrate patterning on LC anchoring.
The results presented in Section \ref{SQPS} were obtained for systems of 864 $\kappa = 3$ HGO particles
confined between two square patterned substrates.
The substrates were separated by a distance $L_{z}$ =
4$\kappa\sigma_{0}$, periodic boundary conditions being imposed
in the $x$- and $y$-directions.

\begin{figure}[!h]
\begin{center}
\includegraphics[scale=0.6]{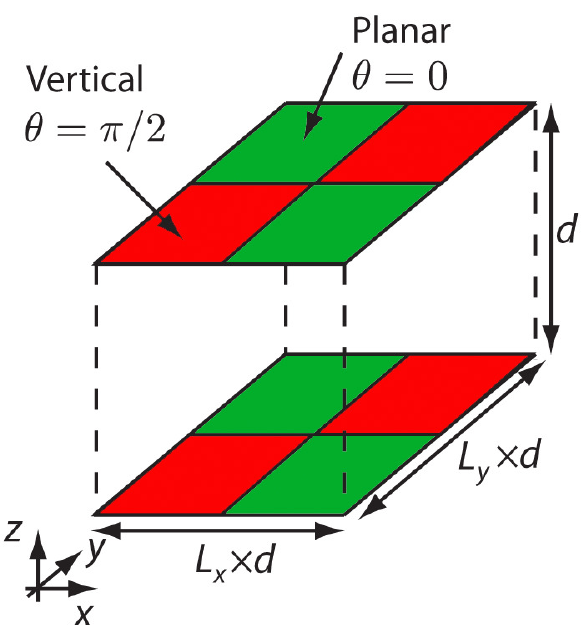}
\caption{(Color online)Schematic representation of rectangle
patterned systems with alternating homeotropic-inducing (red/dark) and planar-inducing (green/light) substrate regions. The Euler angle $\phi$ is 0 from the $x$-axis.} \label{schema_systems}
\end{center}
\end{figure}

On each substrate, $k_{s}$ was set to a
homeotropic-aligning value ($k_{s}=0$) for two quadrants of its area and a planar
value ($k_{s}=3$) for the remainder. Except where otherwise stated,
sharp boundaries have been imposed between the different alignment
regions. The patterns on the top and bottom surfaces
have been kept in perfect registry with one another, as shown
in the schematic in Fig.~\ref{schema_systems}. Each system was
 at low density and compressed, in small increments, by
decreasing the box dimensions $L_{x}$ and $L_{y}$ whilst maintaining the condition $L_{x}$/$L_{y} = 1$. At each density, a run length of 1 million MC sweeps (where one sweep represents one
attempted move per particle) was performed, averages and profiles
being accumulated for the final 500 000 sweeps.

Analysis was performed by dividing stored system
configurations into 100 equidistant constant-$z$ slices and
calculating averages of relevant observables in each slice. This
yielded profiles of quantities such as number density, $\rho^{*}(z)$, from which
structural changes could be assessed. Orientational order profiles
were also measured, particularly
\begin{eqnarray}
Q_{zz}(z)= \frac{1}{N(z)} \sum_{i=1}^{N(z)}\bigg (\frac{3}{2}
u_{i,z}^{2}-\frac{1}{2}\bigg )
\end{eqnarray}
which measures variation across the confined films of orientational
order measured with respect to the substrate normal. Here $N(z)$ is
the instantaneous occupancy of the relevant slice. We have also further
subdivided the system to assess
lateral inhomogeneities induced by the patterning.

\section{Monte Carlo Simulation Results}\label{SQPS}
\begin{figure}[!h]
\subfigure[\label{fig:square_rho30}$\rho*$=0.3
]{\includegraphics[scale=0.106] {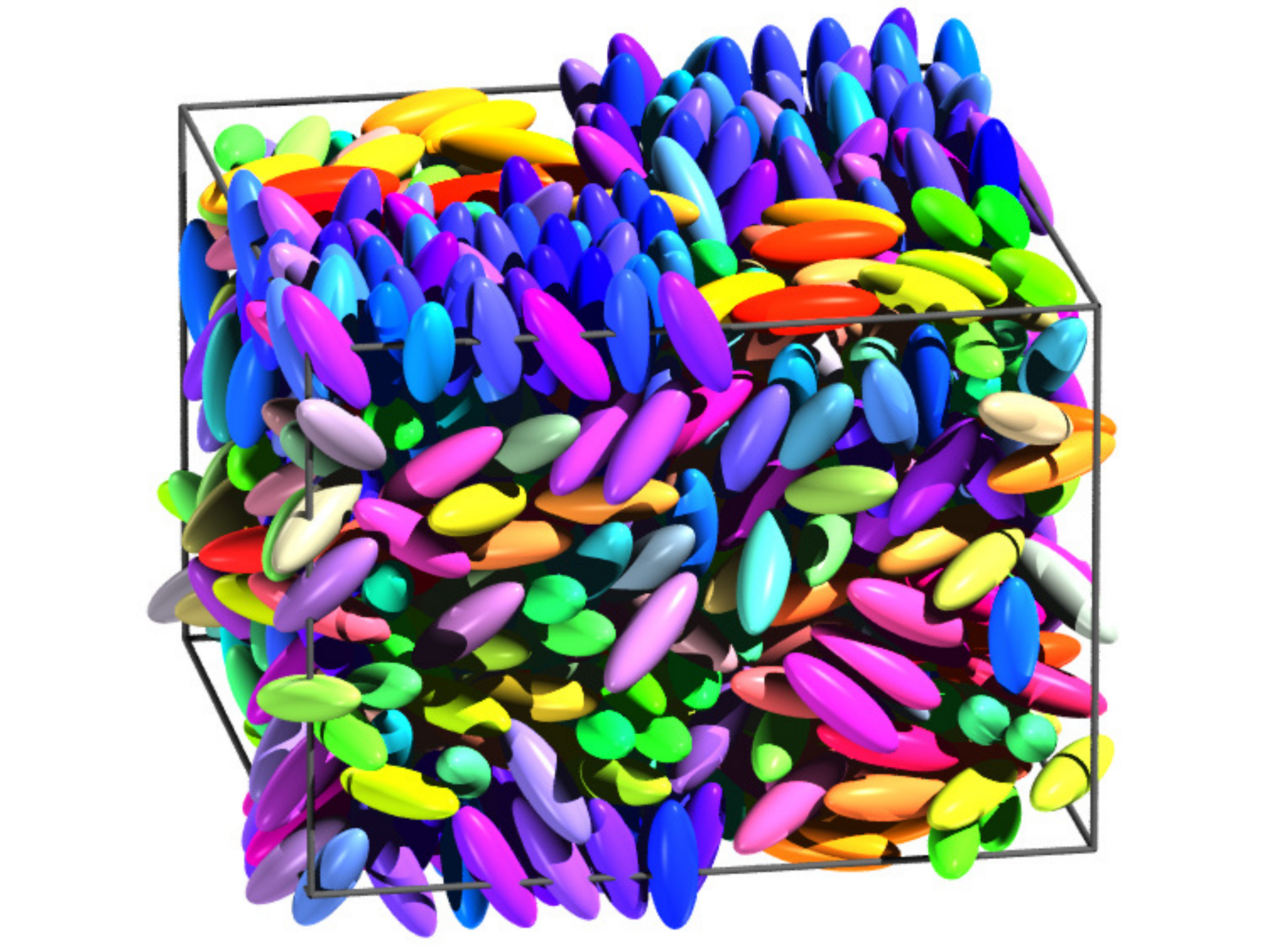}}
\subfigure[\label{fig:square_rho34}$\rho*$=0.34
]{\includegraphics[scale=0.106] {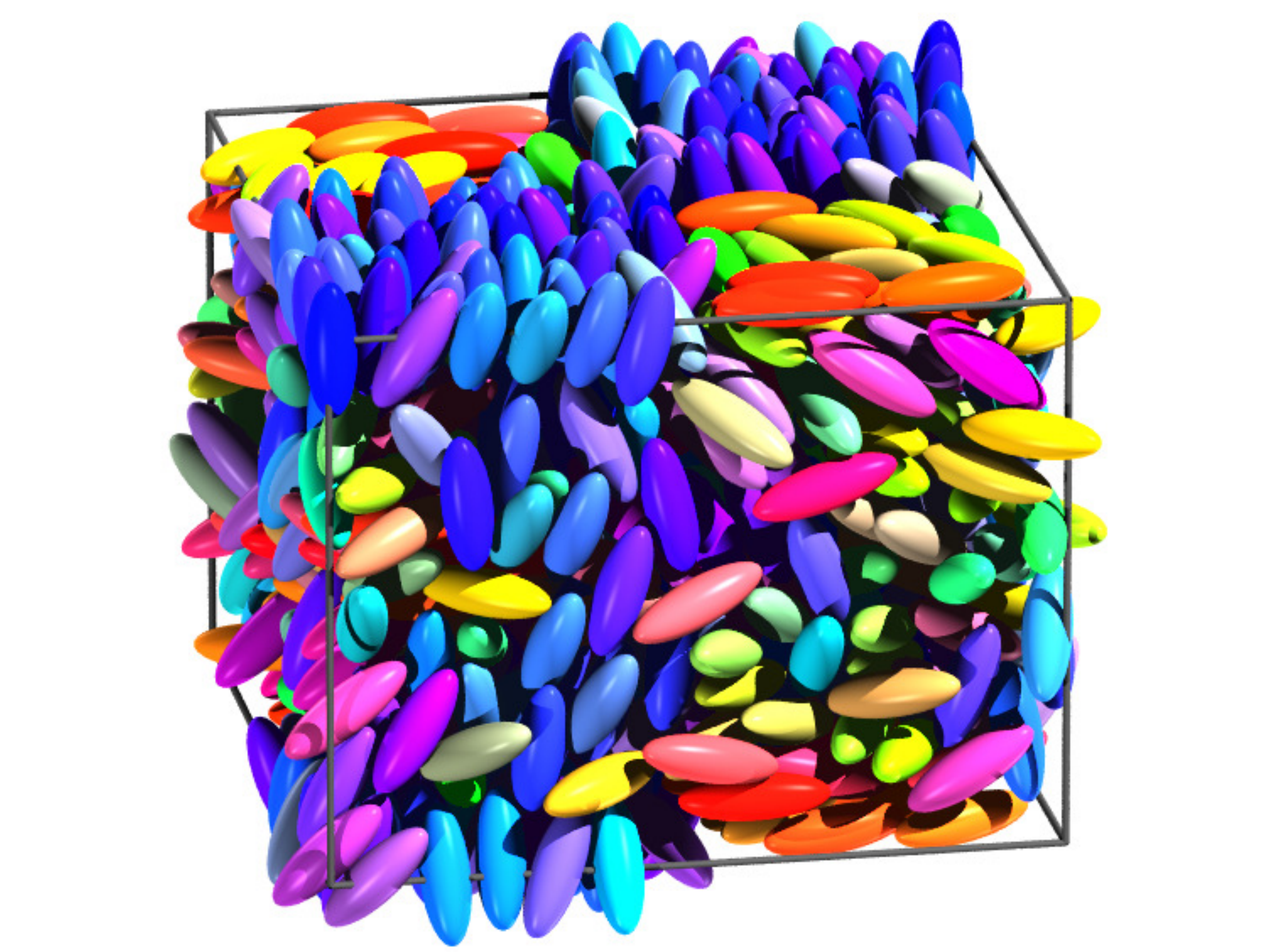}}
\subfigure[\label{fig:square_rho37}$\rho*$=0.37
]{\includegraphics[scale=0.106] {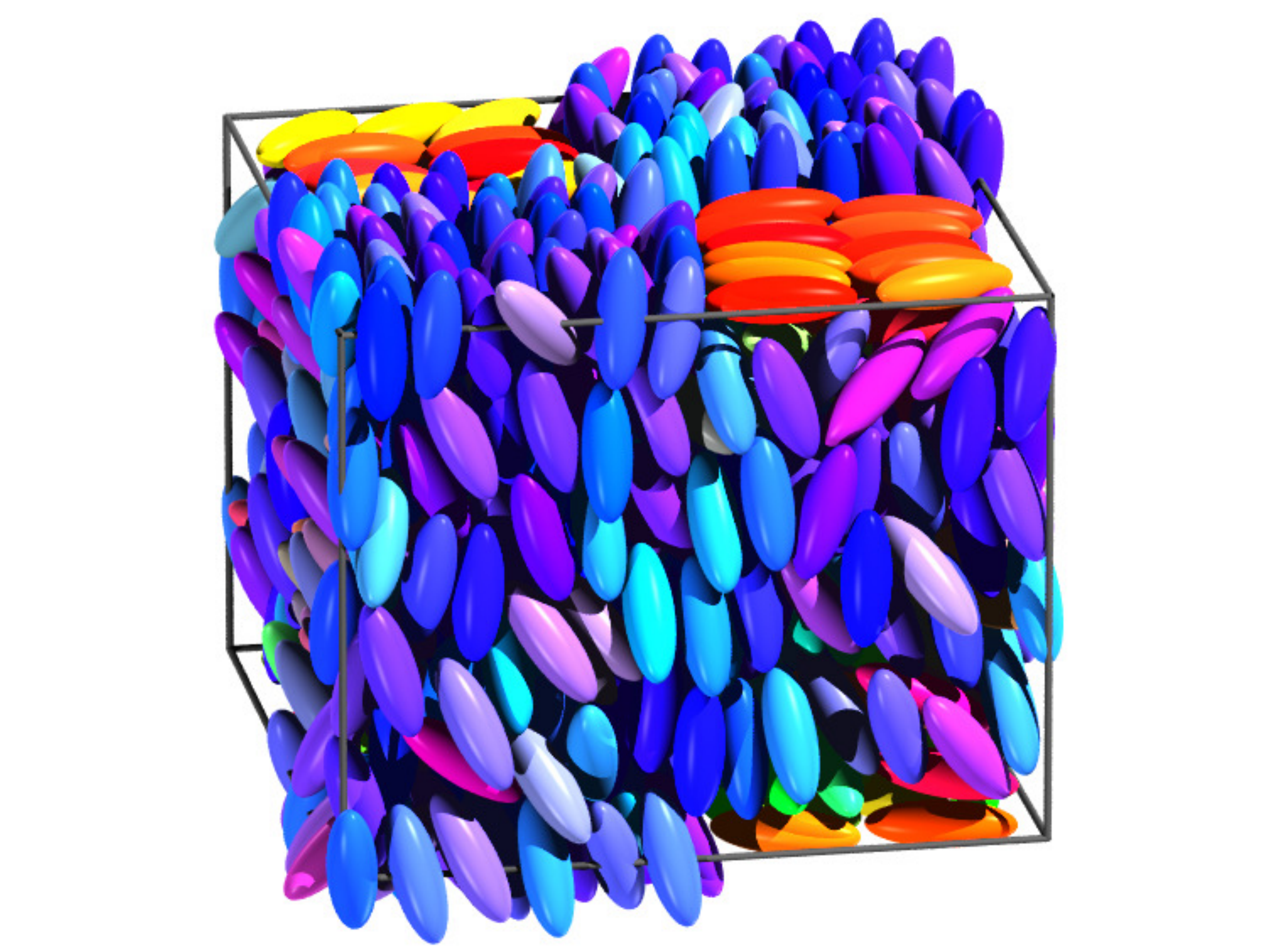}}
\subfigure[\label{fig:square_rho38}$\rho*$=0.38
]{\includegraphics[scale=0.106] {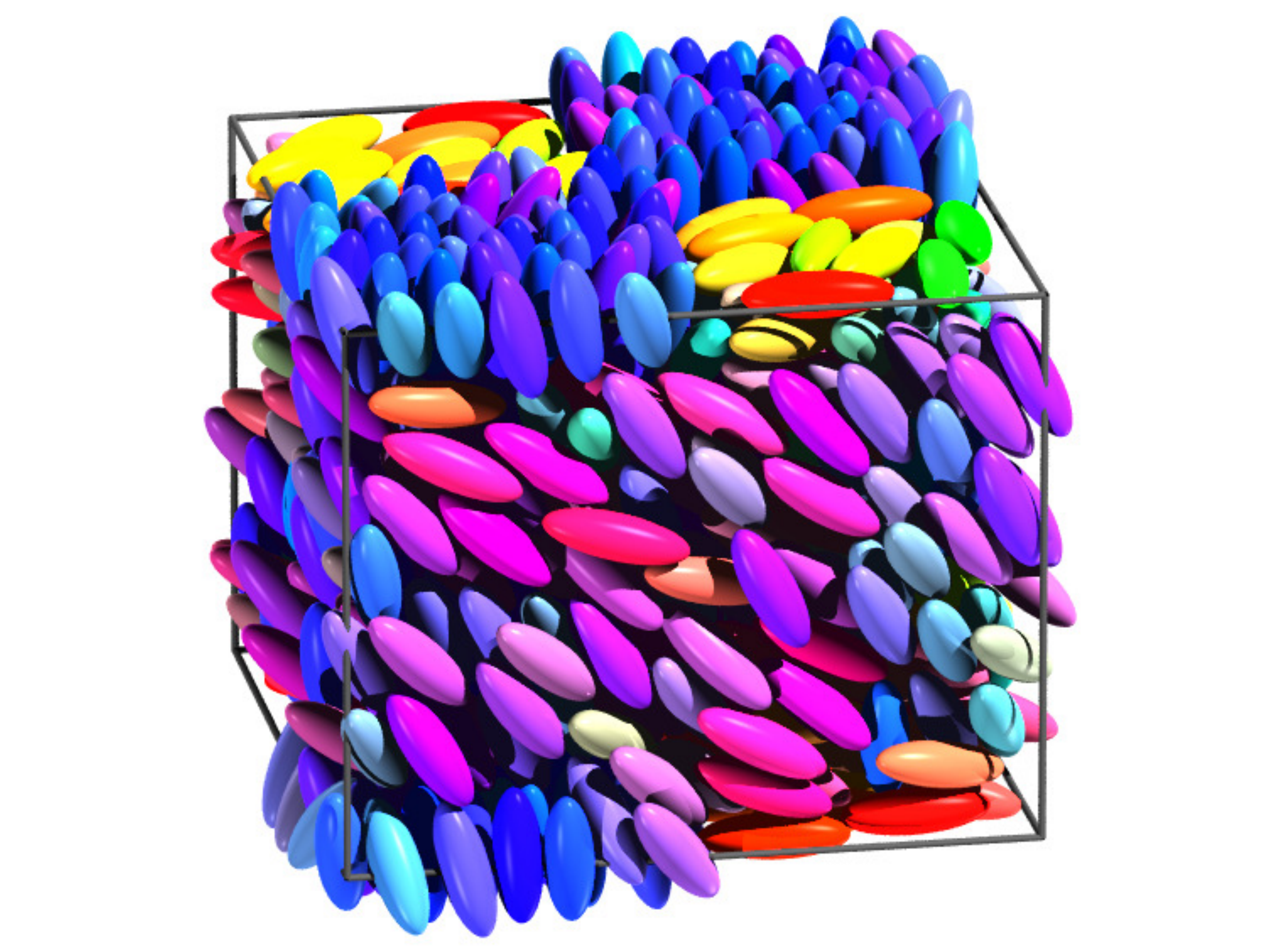}}
\subfigure[\label{fig:square_rho40}$\rho*$=0.40
]{\includegraphics[scale=0.106] {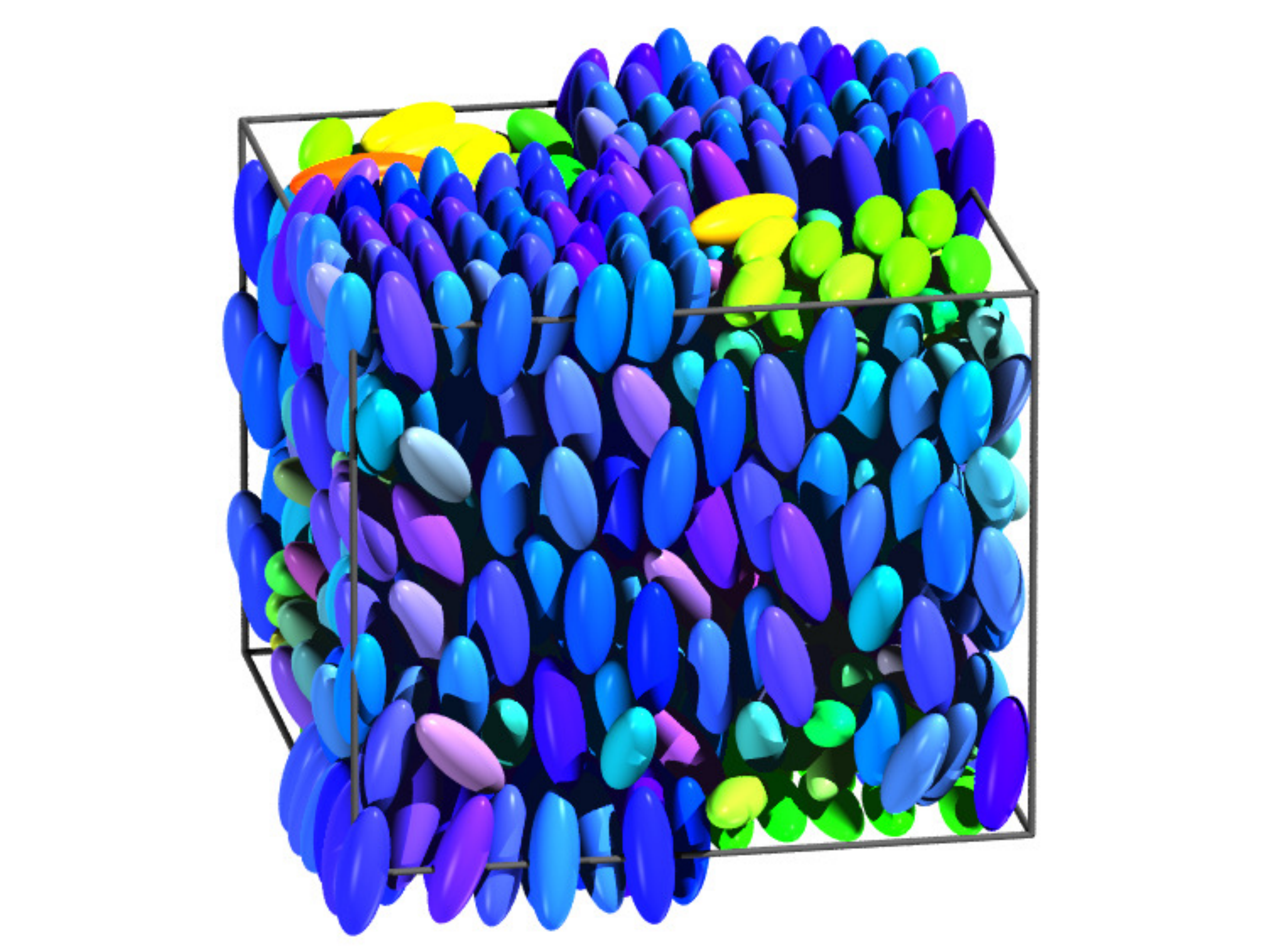}}
\caption{\label{snapshot_square_Lz=4K}(Color online)Snapshots of the square
patterned system with sharp transitions between $k_{s}$=0 and
$k_{s}$=3 regions for a series of different reduced densities. Particles are 0 coded for orientation: $x$ red; $y$ green; $z$ blue.}
\end{figure}

In the initial system considered here, sharp transitions were imposed between the $k_{s}=0$ and $k_{s}=3$ regions. The outcomes of these squared system simulation are 0 by the snapshots shown in Fig.~\ref{snapshot_square_Lz=4K}. Several remarks emerge from these.
The substrate patterning is readily apparent from all of these, with ordered layers of homeotropic-aligned and planar-aligned particles residing in the appropriate regions. Sharp delineation between these regions can be seen for all densities. At $\rho^*$=0.30 and 0.34, the particles at the 0 of the film appear to be relatively disordered, whereas aligned monodomains can be seen at the three higher densities. Animations of these simulations show that in the planar aligning substrate regions, the molecules regularly flipped en masse between the $x$- and $y$- orientations. This tendency is apparent by comparing Figs.~\ref{fig:square_rho37},~\ref{fig:square_rho38} and~\ref{fig:square_rho40}; the orientations on the planar part of the substrate vary from image to image. At high density ($\rho^*$=0.40), the planar parts of the surface patterns appear restricted to the immediate vicinity of the substrate regions. Unlike the equivalent stripe patterned system\cite{RefWorks:24}, for which the patterning fully bridges a film of this width, the planar parts of the pattern appear unable to penetrate the LC film. With this very short length-scale pattern, then, the spontaneous flipping of the planar substrate domains between the two degenerate arrangements appears to make it impossible to fix or control the azimuthal anchoring orientation. At $\rho^*$=0.37, the system appears to exhibit homeotropic anchoring (Fig. ~\ref{fig:square_rho37}). On increasing the density($\rho^*$=0.38), however, this initial homeotropic anchoring adopts a clear tilt (Fig. ~\ref{fig:square_rho38}) through which the planar alignment regions on the two substrates become coupled. On further compression to $\rho^*$=0.40, the bulk director partially regains its alignment normal to the substrates (Fig. ~\ref{fig:square_rho40}).

In the light of these observations, we have 0 the 0 of this system more quantitatively by calculating two sets of profiles of key observables for this system; for analysis purposes, each simulated system has been split in two according to the imposed substrate pattern. In this, individual particles have been allocated to homeotropic-confined or homogeneous-confined regions according to their $x$- and $y$-coordinates.
\begin{figure}[!h]
\begin{center}
\subfigure[\label{fig:homeosquarerho}Density profile: homeotropic-confined region
]{\includegraphics[width=0.5\textwidth] {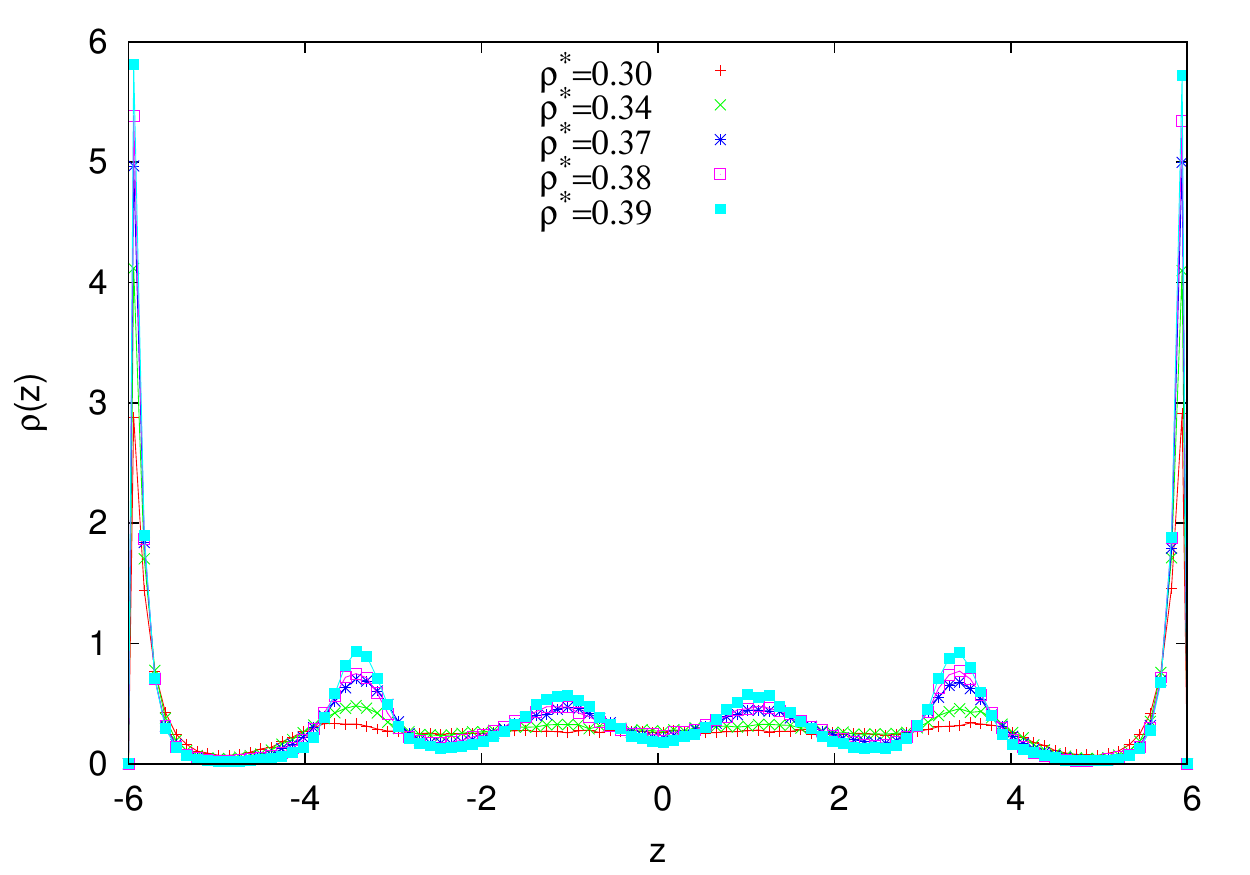}}
\subfigure[\label{fig:planarsquarerho}Density profile: planar-confined region
]{\includegraphics[width=0.5\textwidth] {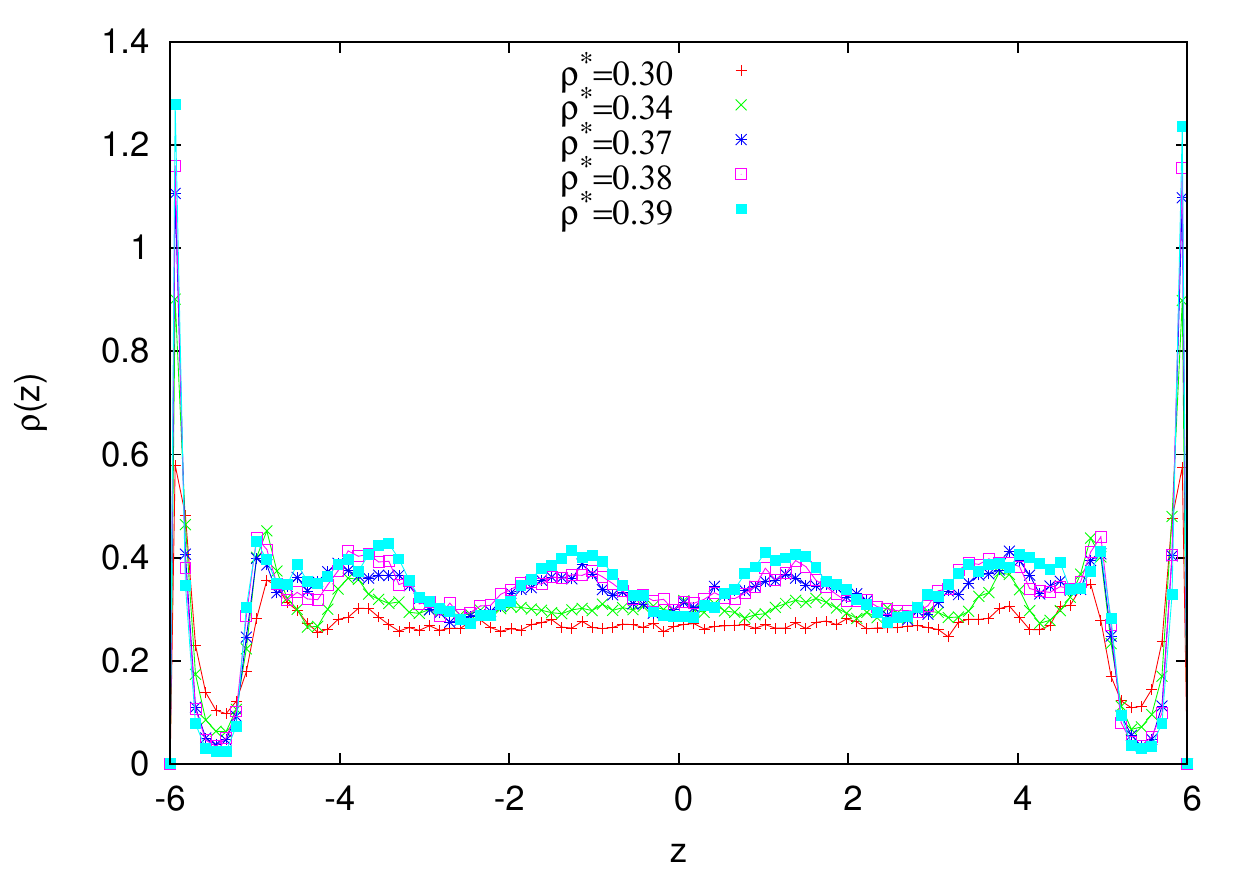}}
\caption{\label{square_Lz=4Krho}(Color online)Density profiles for the square patterned system at different reduced 1 $\rho^*$. }
\end{center}
\end{figure}
 
The density profiles depicted in Fig.~\ref{fig:homeosquarerho} show the adsorption characteristics for the portion of the film confined between the homeotropic surface regions. These indicate that increasing the density leads to formation of surface layers with a periodicity of $\simeq$2$\sigma_{0}$ (i.e. 2/3 of the particle length). Fig. ~\ref{fig:planarsquarerho} shows the corresponding 0 of the regions of the film confined between the planar-
confining surfaces. Here, a shorter wavelength density oscillation is apparent close to the substrates. Despite these differences close to the substrates, both profiles adopt very similar 0 in the central part of the film: essentially featureless at low (isotropic) densities and oscillatory at high (nematic) densities. These oscillations are consistent with the formation of a homeotropic (or near 0) bulk monodomain. Such 0 formation is only seen for much thicker films when stripe patterning is imposed~\cite{RefWorks:24}.

\begin{figure}[!h]
\begin{center}
\subfigure[\label{fig:homeosquareQzz}$Q_{zz}$ profile for the homeotropic-confined region
]{\includegraphics[width=0.5\textwidth]
{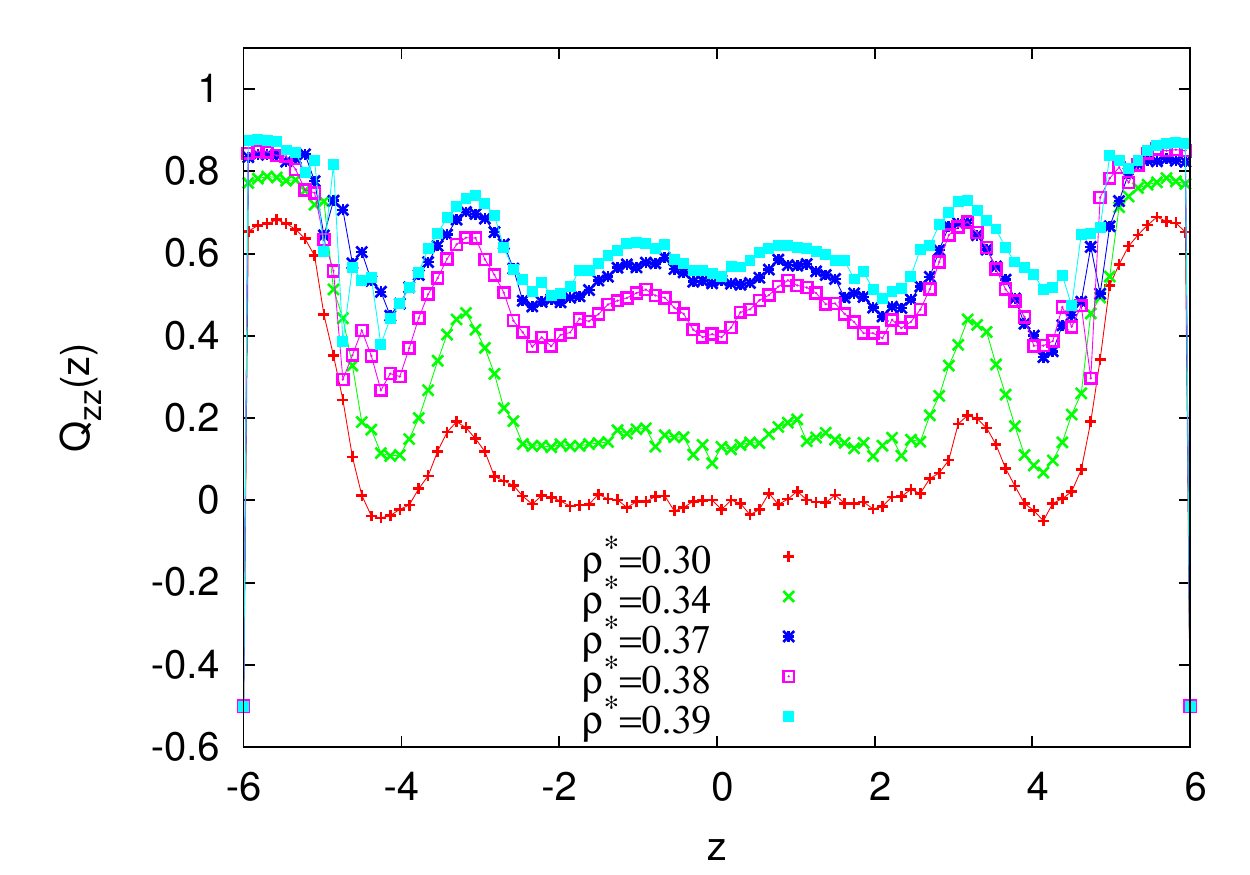}}
\subfigure[\label{fig:planarsquareQzz}$Q_{zz}$ profile for the planar-confined region
]{\includegraphics[width=0.5\textwidth]
{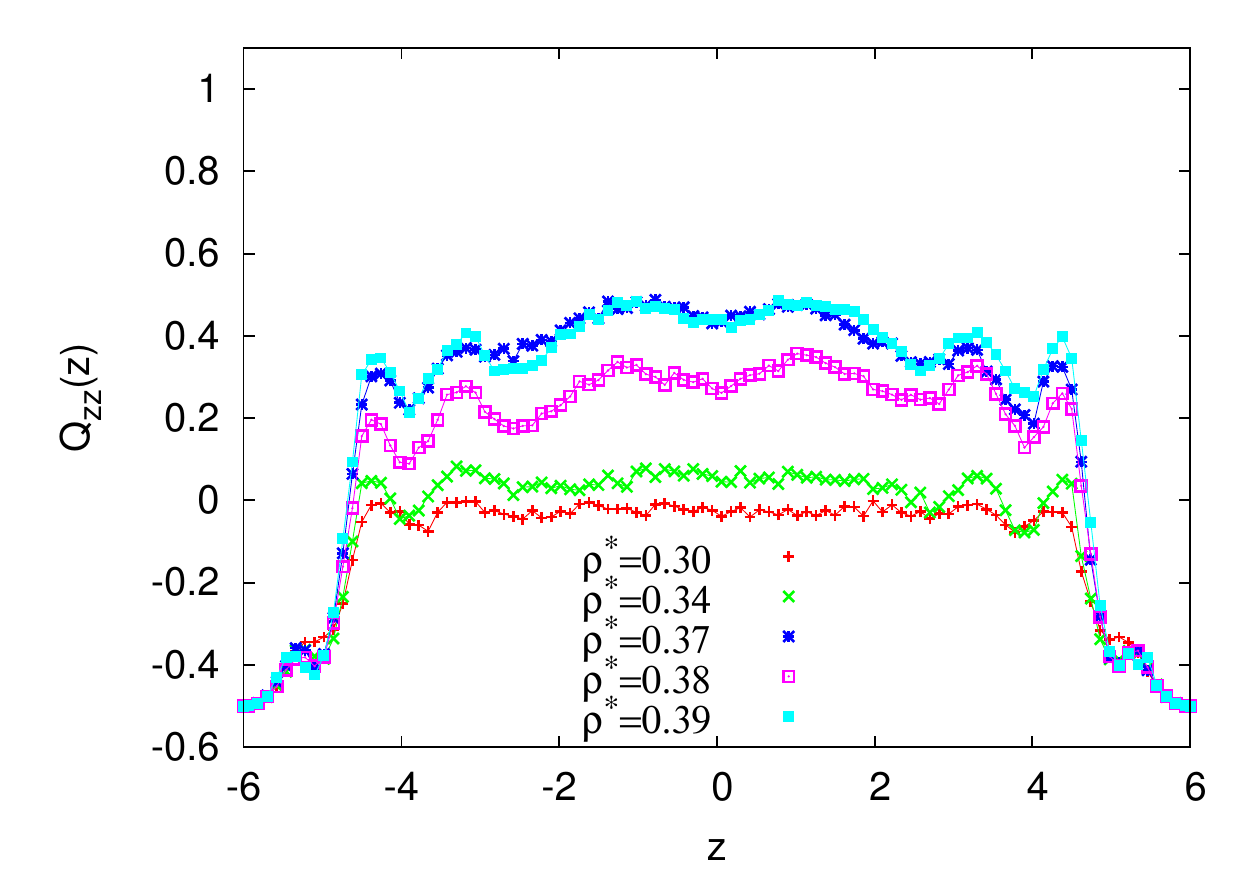}}
\caption{\label{square_Lz=4Kqzz}(Color online)$Q_{zz}$ profiles for the square patterned system at different reduced 1 $\rho^*$.}
\end{center}
\end{figure}

\begin{figure}[!h]
\begin{center}
\includegraphics[width=0.5\textwidth]{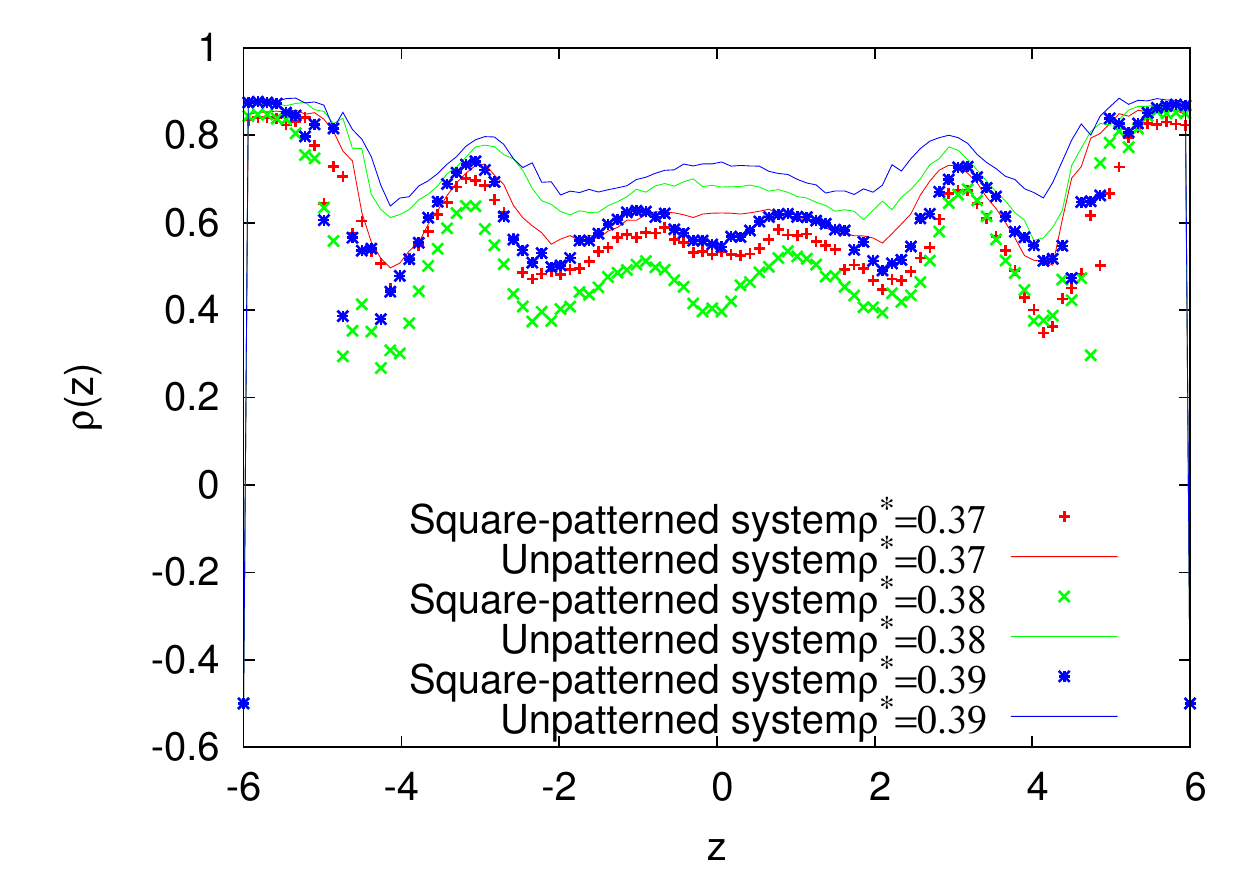}
\caption{\label{fig:comphomeozoom}(Color online)Comparison of the $Q_{zz}$
profiles for unpatterned (lines) and square-patterned (symbols) systems: homeotropic regions.}
\end{center}
\end{figure}

A more complete understanding of the orientational aspects of the substrate-induced ordering in this system can be obtained from the $Q_{zz}$ diagonal component of the order tensor. For perfect homeotropic anchoring, $Q_{zz}(z)$ should tend to 1 and for perfect planar anchoring, $Q_{zz}(z)$ should tend to
-0.5. Fig.~\ref{fig:homeosquareQzz} shows the $Q_{zz}$ profiles
measured in the homeotropic-confined regions. As the density is increased, initially the bulk-region $Q_{zz}$
value increases as well, showing the development of homeotropic
anchoring in the bulk. At a density of 0.37, the
bulk $Q_{zz}$ value reaches 0.60-0.65. On further increasing
the density to 0.38, however, the $Q_{zz}$ value decreases to just
below 0.5. Then, as the density reaches 0.4, the $Q_{zz}$
increases again to $Q_{zz}$=0.60-0.65. This non-monotonic 0 confirms, in a statistically significant fashion, the tilt 0 apparent in the corresponding snapshots.
\newline
It is also informative to compare these observations with equivalent profiles obtained for HGO films confined between unpatterned homeotropic- and planar-aligning substrates. To this end, Fig.~\ref{fig:comphomeozoom} shows that for
the equivalent unpatterned homeotropic-aligning system, increasing the density causes the central $Q_{zz}$ value to increase monotonically(red line-green line-dark blue line/gray line-light gray line-dark gray line). The fact that $Q_{zz}$ shows a decreases at $\rho^*\simeq$0.37 in the patterned system is, then,
associated with a tilt of the bulk director caused by the planar pattern regions on the surface.

An equivalent comparison performed for the planar-aligning region (Fig.~\ref{fig:compplanar})shows a marked difference between the patterned and unpatterned systems. Indeed, despite its intrinsic anchoring 0, the $Q_{zz}$(z) 0 of the  planar-aligned region of the patterned system is actually far closer to that of the unpatterned homeotropic-confined system. Only very close to the substrates is the planar nature of the imposed substrate pattern apparent.

\begin{figure}[!h]
\begin{center}
\includegraphics[width=0.5\textwidth]{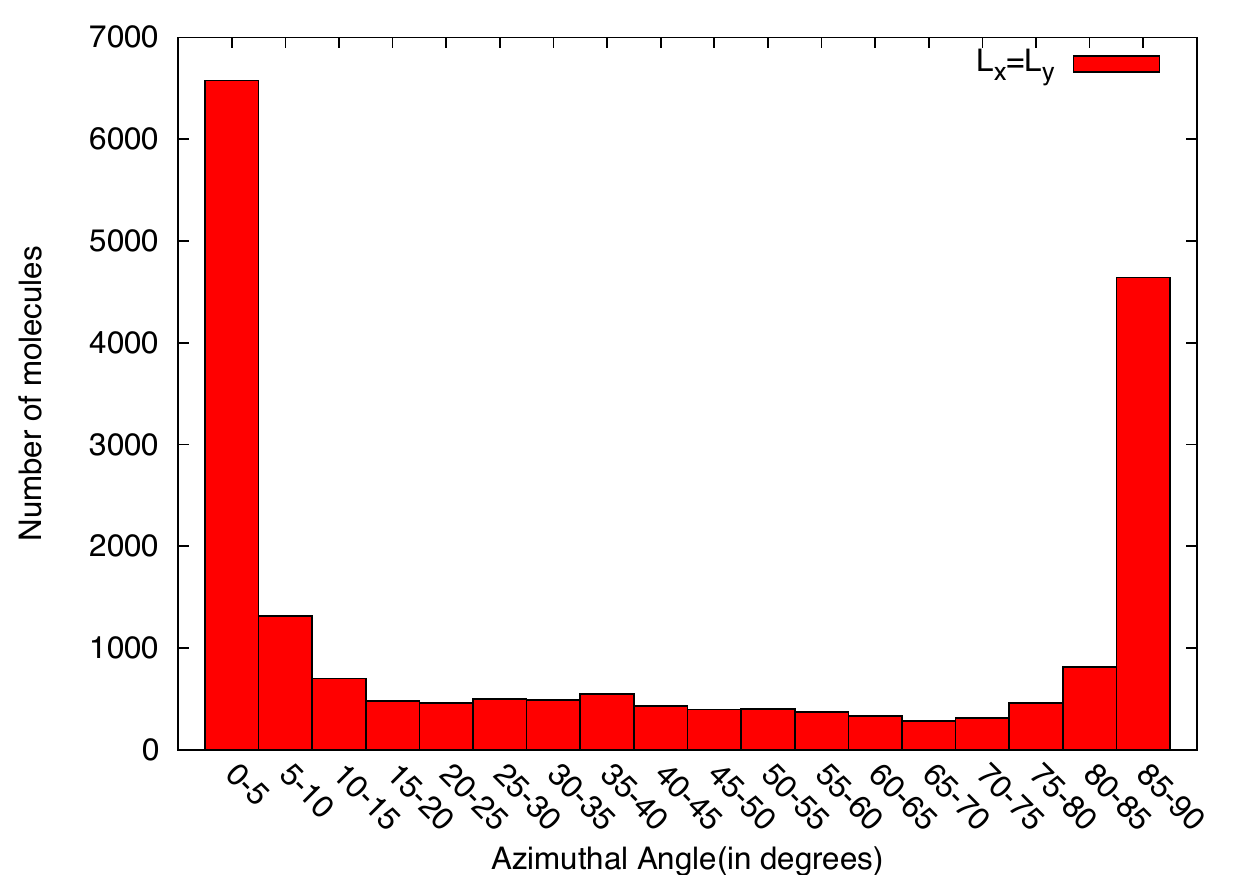}
\caption{\label{histo_square}(Color online)Histogram showing the azimuthal
angle distribution of surface particles in the P region.}
\end{center}
\end{figure}

\begin{figure}[!h]
\begin{center}
\includegraphics[width=0.5\textwidth]{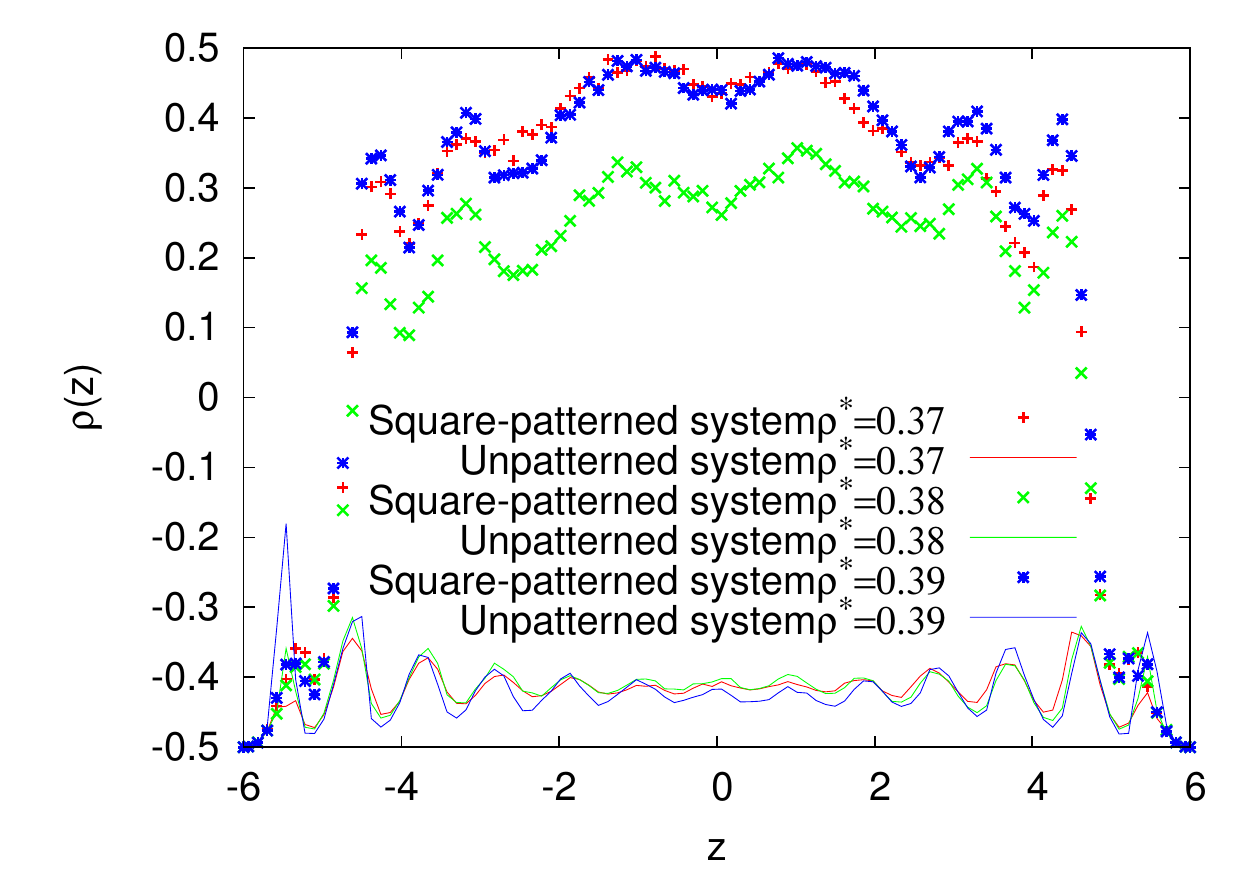}
\caption{\label{fig:compplanar}(Color online)Comparison of the $Q_{zz}$
profiles for unpatterned (lines) and square-patterned (symbols) systems: planar regions.}
\end{center}
\end{figure}

In order to assess the azimuthal anchoring 0 in this system, we also constructed a time-averaged histogram of the molecular 0 angles observed during the $\rho^*$=0.37 simulation. Specifically, this histogram (Fig .~\ref{histo_square}) was generated from 500 stored configuration files and based on the orientations of particles
within 1$\sigma_{0}$ of the planar substrate regions. The histogram is strongly peaked at angles corresponding to the boundaries of the square pattern, i.e. the molecules at the planar substrates are strongly
disposed to adopting azimuthal angles $\phi$ of $\simeq 0^\circ\leq\phi\leq5^\circ$ and $\simeq 90^\circ\leq\phi\leq90^\circ$. This is consistent with our previous observation that the molecules on this region appeared to regularly flip between the $x$- and $y$-directions.

\begin{figure}[!h]
\begin{center}
\includegraphics[width=0.5\textwidth]{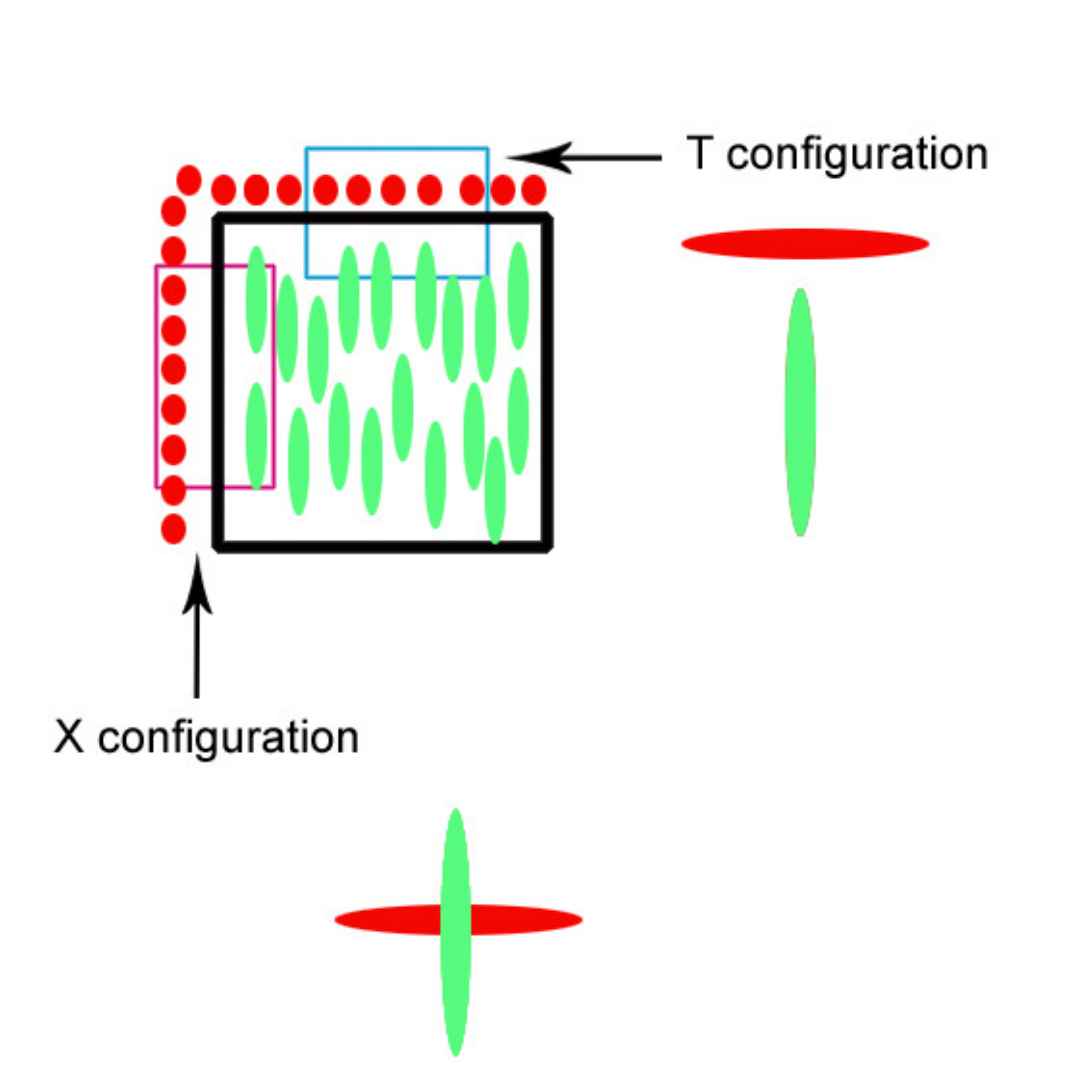}
\caption{\label{Boundaries_green}(Color online)Schematic x-y plane slice through
a hypothetical planar aligned bridging domain in a square
patterned system.T-configuration : splay/bend distortion. X-configuration : Twist }
\end{center}
\end{figure}

Before closing this section, we return to the observation that the substrate patterning applied here failed to penetrate the LC film beyond the first adsorbed monolayer. In fact, we can report that this was a general characteristic observed for a range of different two-dimensional patternings; simulations performed with circle, oval and rectangle patterns, respectively, all led to the development of central monodomain configurations. To explain why the orientational bridging observed in stripe-patterned systems is lost on moving to two-dimensional patterns, we consider the hypothetical square bridging domain shown schematically in Fig.~\ref{Boundaries_green}. Here a planar-aligned domain is bounded at each face by homeotropic material. From the schematic, though, it is clear that two distinct pairs of domain boundaries are required for this: one pair involving T-like orientational changes and another involving X-like configurations. Such a scenario is clearly unstable: the surface tensions associated with these different pairs of interfaces could not, in general, be equal. As a consequence, the hypothetical square bridging domain considered here could never be a stable arrangement. Indeed, similar instabilities hold for all orientational bridges projected from two-dimensional patternings.

To conclude, these simulations indicate that LC 0 between square-patterned substrates have a tendency to form monodomains. These monodomains are different from those developed between unpatterned substrates, though, since a) they can exhibit a non-monotonic density-dependent tilt and b) the azimuthal anchoring shows a strong coupling parallel to the square edges but is degenerate between the different edge orientations.
\section{Continuum model}
To further understand the aligning effects observed in the simulations
presented in the previous section, we now consider the 0 of systems 
with the same geometry of patterning but applied at a much larger length-scale. Specifically, 
we analyze the effect of square-patterned substrates on LC films in the continuum
limit. In this approach, the local orientation of the nematic is characterized
by a unit vector field known as the director and 0 here
by
\begin{equation}
\mathbf{\hat{n}}(\vec{x})=\left(\cos\theta\sin\phi,\cos\theta\cos\phi,\sin\theta\right)\label{eq:director}
\end{equation}
where the coordinates are chosen as depicted in fig. 2. The actual
configuration adopted by the nematic is that which minimizes the Frank
free energy
\begin{widetext}
\begin{eqnarray}
F=\frac{1}{2}\int d^{3}x\ K_{1}\left(\nabla\cdot\mathbf{\hat{n}}\right)^{2}+K_{2}\left[\mathbf{\hat{n}}\cdot(\nabla\times\mathbf{\hat{n}})\right]^{2}+K_{3}
\left|\mathbf{\hat{n}}\times(\nabla\times\mathbf{\hat{n}})\right|^{2}+\int_{s}\text{d}S\ g(\vec{\mathbf{n}},\vec{\mathbf{n}}_{0}).
\end{eqnarray}
\end{widetext}
Here, the first integral is to be performed over the volume of the nematic
layer and the second over the surfaces in contact with the 0.
The interaction of the nematic with the surface is characterized by
an anchoring potential $g(\vec{\mathbf{n}},\vec{\mathbf{n}}_{0})$
that measures the energy cost of moving the director away from an
easy axis $\vec{\mathbf{n}}_{0}$; for a patterned surface this varies
as a function of position. The configuration of the LC
is found by solving the Euler{-}Lagrange equations for $\theta$ and
$\phi$; these are generally nonlinear and difficult to solve analytically
in more than one spatial dimension.

A common simplification, known as the one-constant approximation,
is to set $K_{1}=K_{2}=K_{3}$. If this is done, the Euler{-}Lagrange
equation for $\theta$ reduces to Laplace's equation. However, such
an approximation is unsuitable for analyzing situations with patterned
surfaces because the aligning effect on the LC is due
to \emph{differences} between the elastic constants\cite{RefWorks:11}. A ``two-constant''
approximation, where $K_{1}=K_{3}\neq K_{2}$ has been previously
used to understand the situation of a nematic film in contact with
a surface patterned with alternating homeotropic and planar stripes,
a two-dimensional system\cite{RefWorks:11,RefWorks:12} and here we extend the analysis
to three dimensions.

In order to proceed, a further simplifying assumption is made: that
the director is confined everywhere to a single plane, i.e. that $\phi$
is spatially uniform. This simplification is motivated (and justified) by the 
observation that the molecular distribution of azimuthal angles (Fig .~\ref{histo_square}) in 
our MC simulations implies a monodomain arrangement for all nematic films. Whilst $\phi$
is taken to be constant, the zenithal, or tilt, angle $\theta$ remains free to vary in response to the substrate pattern. Keeping$phi$fixed is further motivated by the observation that wherever the director
is nearly homeotropic, variations in $\phi$ contribute negligibly
to the free energy. The free energy density in this situation is
\begin{widetext}
\begin{eqnarray}
f & = & \frac{1}{2}\left\{ \left(\tau\cos^{2}\phi+\sin^{2}\phi\right)\left(\frac{\partial\theta}{\partial x}\right)^{2}+\left(\tau\sin^{2}\phi+\cos^{2}\phi
\right)\left(\frac{\partial\theta}{\partial y}\right)^{2}+\right.\nonumber \\
 &  & \left.\ \ \ \ \ +(1-\tau)\sin(2\phi)\frac{\partial\theta}{\partial x}\frac{\partial\theta}{\partial y}+\left(\frac{\partial\theta}{\partial z}\right)^
{2}\right\} \label{eq:FreeEnergy}
\end{eqnarray}
\end{widetext}
and the corresponding Euler\-Lagrange equation for $\theta$ is linear
\begin{widetext}
\begin{eqnarray}
\left(\tau\cos^{2}\phi+\sin^{2}\phi\right)\frac{\partial^{2}\theta}{\partial x^{2}}+\left(\tau\sin^{2}\phi+\cos^{2}\phi\right)\frac{\partial^{2}\theta}
{\partial y^{2}}+(1-\tau)\sin(2\phi)\frac{\partial^{2}\theta}{\partial x\partial y}+\frac{\partial^{2}\theta}{\partial z^{2}}=0.\label{eq:EulerLangrange}
\end{eqnarray}
\end{widetext}
This can be converted to Laplace's equation in new coordinates $(\xi,\eta,\zeta)$
by the following linear transformation
\begin{equation}
\left(\begin{array}{c}
\xi\\
\eta\\
\zeta
\end{array}\right)=P^{T}QP\left(\begin{array}{c}
x\\
y\\
z
\end{array}\right)\label{eq:mapping}
\end{equation}
where
\begin{equation}
P=\left(\begin{array}{ccc}
\cos\left(\phi+\frac{\pi}{4}\right) & -\sin\left(\phi+\frac{\pi}{4}\right) & 0\\
\sin\left(\phi+\frac{\pi}{4}\right) & \cos\left(\phi+\frac{\pi}{4}\right) & 0\\
0 & 0 & 1
\end{array}\right)\label{eq:P}
\end{equation}
and
\begin{equation}
Q=\left(\begin{array}{ccc}
\frac{1}{2}\left(1+\frac{1}{\sqrt{\tau}}\right) & \frac{1}{2}\left(\frac{1}{\sqrt{\tau}}-1\right) & 0\\
\frac{1}{2}\left(\frac{1}{\sqrt{\tau}}-1\right) & \frac{1}{2}\left(1+\frac{1}{\sqrt{\tau}}\right) & 0\\
0 & 0 & 1
\end{array}\right).\label{eq:Q}
\end{equation}
The geometric interpretation of the transformation is a combination
of a rotation and shear. To solve the Euler-Lagrange equation for
$\theta(x,y,z)$, eq. \ref{eq:EulerLangrange} we try a solution of
the form
\begin{widetext}
\begin{eqnarray}
\theta(x,y,z)=\theta_{0}+\sum_{n=-\infty}^{\infty}\sum_{m=-\infty}^{\infty}\frac{1}{\lambda}\left(A_{nm}e^{-\nu_{nm}z}+B_{nm}e^{\nu_{nm}z}\right)\exp\left[i2
\pi\left(nx+my\right)/\lambda\right]\label{eq:solution}
\end{eqnarray}
\end{widetext}
where $\lambda=L_x / L_z$ such that $\lambda L_z$ is the period the patterning in both $x$ and $y$
directions. The equation is satisfied if the parameters $\nu_{nm}$
are chosen thus

\begin{widetext}
\begin{eqnarray}
\nu_{nm}=\pi\sqrt{2(\tau+1)\left(m^{2}+n^{2}\right)-2(\tau-1)\left[2mn\sin(2\phi)+\cos(2\phi)\left(m^{2}-n^{2}\right)\right]}\label{eq:nu}
\end{eqnarray}
\end{widetext}
The constant $\theta_{0}$ is, from the mean-value theorem,
\begin{equation}
\theta_{0}=\frac{\pi}{4}\label{eq:theta0}
\end{equation}
The coefficients $A_{nm}$ and $B_{nm}$ are determined by the boundary
conditions. For weak anchoring, these are the torque-balance equation
\begin{equation}
\hat{\bf s}\cdot\frac{\partial f(\theta,\nabla\theta)}{\partial\nabla\theta}+\frac{\partial g(\theta,\theta_{e})}{\partial\theta}=0\label{eq:torquebalanceold}
\end{equation}
evaluated at each surface, where $\hat{\bf s}$ is the \emph{outward} surface
normal. To facilitate separation of the 0 in \ref{eq:theta0},
the harmonic anchoring potential
\begin{equation}
g_{H}(\theta,\theta_{e})=\frac{W_{\theta}}{2}(\theta-\theta_{e})^{2}\label{eq:harmonic}
\end{equation}
is chosen, yielding the Robin boundary condition
\begin{equation}
\pm L_{\theta}\frac{\partial\theta}{\partial z}+\theta=\theta_{e}\label{eq:Robin}
\end{equation}
where the $-$ve sign corresponds to $z=z_0=-L_z/2$, the $+$ve sign to
$z=z_0=+L_z/2$ and the dimensionless parameter associated with polar
anchoring $L_{\theta}$ is
\begin{equation}
L_{\theta}=\frac{K_{1}}{W_{\theta}d}.\label{eq:AnchoringLength}
\end{equation}
Inserting the solution (\ref{eq:solution}) into the boundary condition
(\ref{eq:Robin}) at each surface yields the coupled system of equations
\begin{widetext}
\begin{eqnarray}
\left(\begin{array}{cc}
1+L_{\theta}\nu_{nm} & 1-L_{\theta}\nu_{nm}\\
e^{-\nu_{nm}}(1-L_{\theta}\nu_{nm}) & e^{\nu_{nm}}(1+L_{\theta}\nu_{nm})
\end{array}\right)\left(\begin{array}{c}
A_{nm}\\
B_{nm}
\end{array}\right)=\left(\begin{array}{c}
c_{nm}\\
d_{nm}
\end{array}\right)\label{eq:coupledeq}
\end{eqnarray}
\end{widetext}
where $c_{nm}$ and $d_{nm}$ are the Fourier coefficients of the
easy axis profile $\theta_{0}(x,y)$ at the $z=0$ and $z=1$ surfaces
respectively. These are simply
\begin{equation}
c_{nm}=d_{nm}=\begin{cases}
-\frac{\lambda}{\pi nm} & n,m\ \text{odd}\\
0 & \text{otherwise}
\end{cases}\label{eq:cdcoeff}
\end{equation}
and solution of (\ref{eq:coupledeq}) yields
\begin{eqnarray}
A_{nm} & = & \frac{e^{\nu_{nm}}c_{nm}}{L_{\theta}\nu_{nm}(e^{\nu_{nm}}-1)+(e^{\nu_{nm}}+1)},\nonumber \\
B_{nm} & = & \frac{c_{nm}}{L_{\theta}\nu_{nm}(e^{\nu_{nm}}-1)+(e^{\nu_{nm}}+1)}.\label{eq:ABcoeff}
\end{eqnarray}
The complete director profile for given values of $\phi$, $\tau$,
$L_{\theta}$ and $\lambda$ is then fully specified by the series solution
(\ref{eq:solution}) and the parameters (\ref{eq:nu}) and (\ref{eq:ABcoeff})
that have now been determined.

The free energy associated with the solution (\ref{eq:solution})
may be evaluated by substituting it into the free energy (\ref{eq:FreeEnergy})
and performing necessary integrations. The bulk energy is
\begin{widetext}
\begin{eqnarray}
F_{b} & = & \sum_{nm}\frac{\pi^{2}}{\lambda^{2}\nu_{nm}}\left[\left(A_{nm}^{2}e^{-\nu_{nm}}+B_{nm}^{2}e^{+\nu_{nm}}\right)\sinh(\nu_{nm})+2A_{nm}B_{nm}\nu_
{nm}\right]\times\nonumber \\
 &  & \ \ \ \ \ \times\left\{ (1+\tau)\left(m^{2}+n^{2}\right)+(1-\tau)\left[\cos(2\phi)\left(m^{2}-n^{2}\right)+2mn\sin(2\phi)\right]\right\} +\nonumber \\
 &  & +\sum_{nm}\frac{1}{2}\nu_{nm}\left[\left(A_{nm}^{2}e^{-\nu_{nm}}+B_{nm}^{2}e^{+\nu_{nm}}\right)\sinh(\nu_{nm})-2A_{nm}B_{nm}\nu_{nm}\right].\label
{eq:bulkenergy}
\end{eqnarray}
\end{widetext}
The surface energy (for each surface) is
\begin{equation}
F_{s}=\pi^{2}\lambda^{2}/16+\frac{1}{L_{\theta}}\sum_{nm}(A_{nm}+B_{nm})(A_{nm}+B_{nm}-2c_{nm}).\label{eq:surf}
\end{equation}

These expressions for the free energy have been evaluated numerically as
a function of $\phi$ for different values of $L_{\theta}$. A value
of $\tau=K_{2}/K_{1}=1/2$ was used that is approximately valid for
many common nematics including 5CB. The period of the pattern was
initially chosen to be the same as the cell thickness, i.e. $\lambda=1$.
\begin{figure}[!h]
\begin{center}
\includegraphics[width=0.47\textwidth]{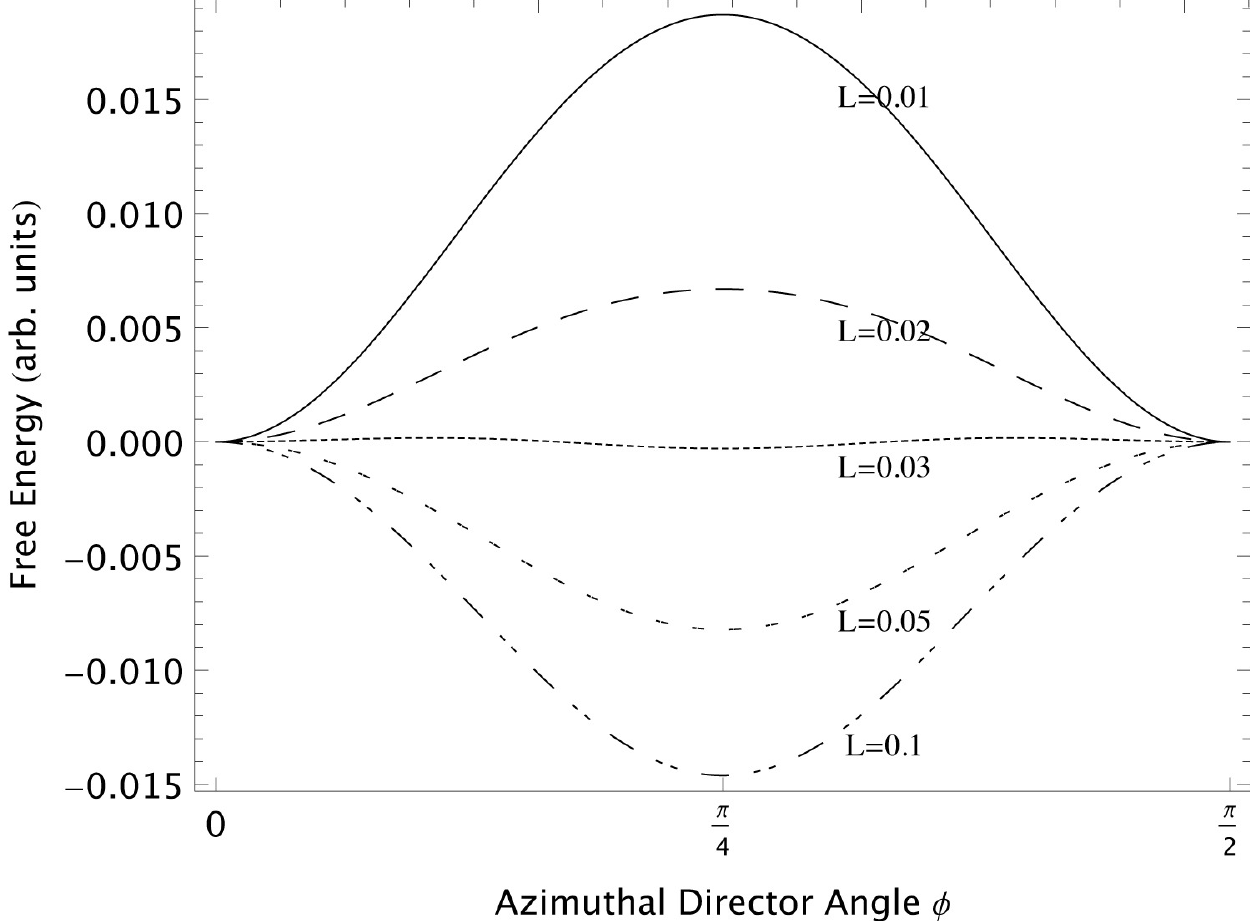}
\caption{\label{fig:AzimuthalVariationFE}Free energy of the nematic as a function
of $\phi$, plotted for various values of $L_{\theta}$.}
\end{center}
\end{figure}
The plots displayed in fig.~\ref{fig:AzimuthalVariationFE} reveal
an anchoring transition: as $L_{\theta}\to0$, representing rigid
polar anchoring, the squares promote azimuthal alignment parallel
to their sides, and there are two degenerate solutions at $\phi=0$
or $\phi=\pi/2$, i.e. the same 0 as was observed in our molecular simulations. 
If $L_{\theta}$ is increased, however,
alignment along the diagonals i.e. $\phi=\pm\pi/4$ becomes the energetically
preferred solution. The critical $L_{\theta}$ at which the diagonal
and aligned solutions become degenerate is roughly $L_{\theta}\sim0.03$.
We can relate this result to experiment by taking $K_{1}\approx6\text{pN}$
(5CB), a typical cell thickness $d=10\mu\text{m}$ and a typical value
for $W_{\theta}=10^{-5}\text{J}\text{m}^{-2}$, which together give $L_{\theta}=0.006$.
This falls well into the regime of $\phi=0$ and $\pi/2$ anchoring which, indeed, 
is precisely what was seen by Bramble~\cite{RefWorks:23}.

\begin{figure}[!h]
\includegraphics[width=0.47\textwidth]{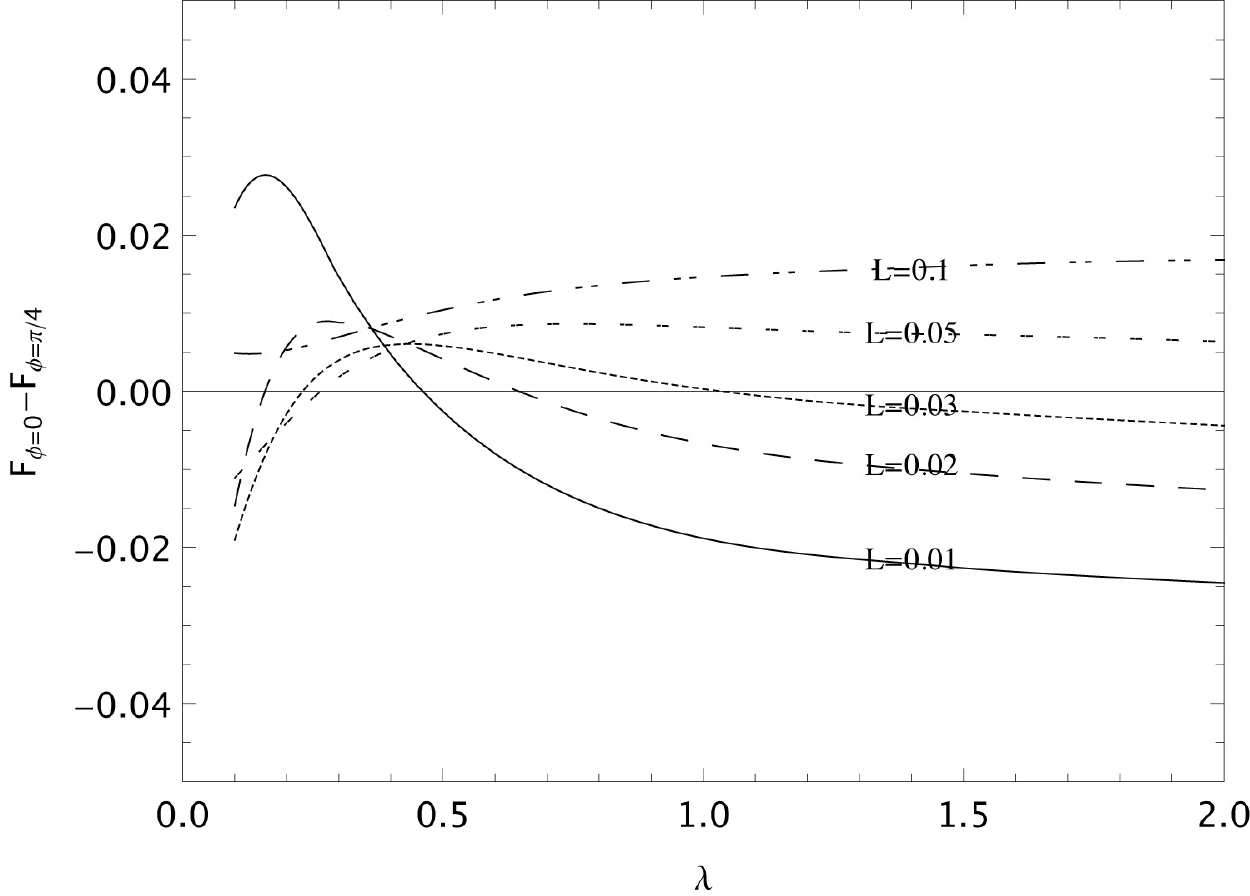}
\caption{\label{fig:FreeEnergyDiff}Free energy difference between the aligned
$\phi=0$ and diagonal $\phi=\pi/4$ solutions as a function of the
period of the pattern $\lambda$. }
\end{figure}

The second parameter of interest is $\lambda$, the overall size of the squares
relative to the cell thickness. Shown in figure \ref{fig:FreeEnergyDiff}
is the energy difference between the aligned $\phi=0$ and diagonal
$\phi=\pi/4$ solutions as a function of $\lambda$, plotted for various
values of $L_{\theta}$. In this plot, therefore, the diagonal solution is stable where lines lie 
below the abscissa, whereas the edge aligned solution is stable where the lines take positive values.
We see, therefore, that the aligned solution is preferred both at small $/lambda$ and 
as $\lambda\to\infty$. Reducing $L_{\theta}$ has the effect of narrowing the
window of $\lambda$ values for which the diagonal solution is preferred. 

For comparison, regarding the Monte Carlo simulation, $\lambda$=1.12 at reduced density of 0.4 (Fig. ~\ref{fig:square_rho40}).

\begin{figure}[!h]
\includegraphics[width=0.47\textwidth]{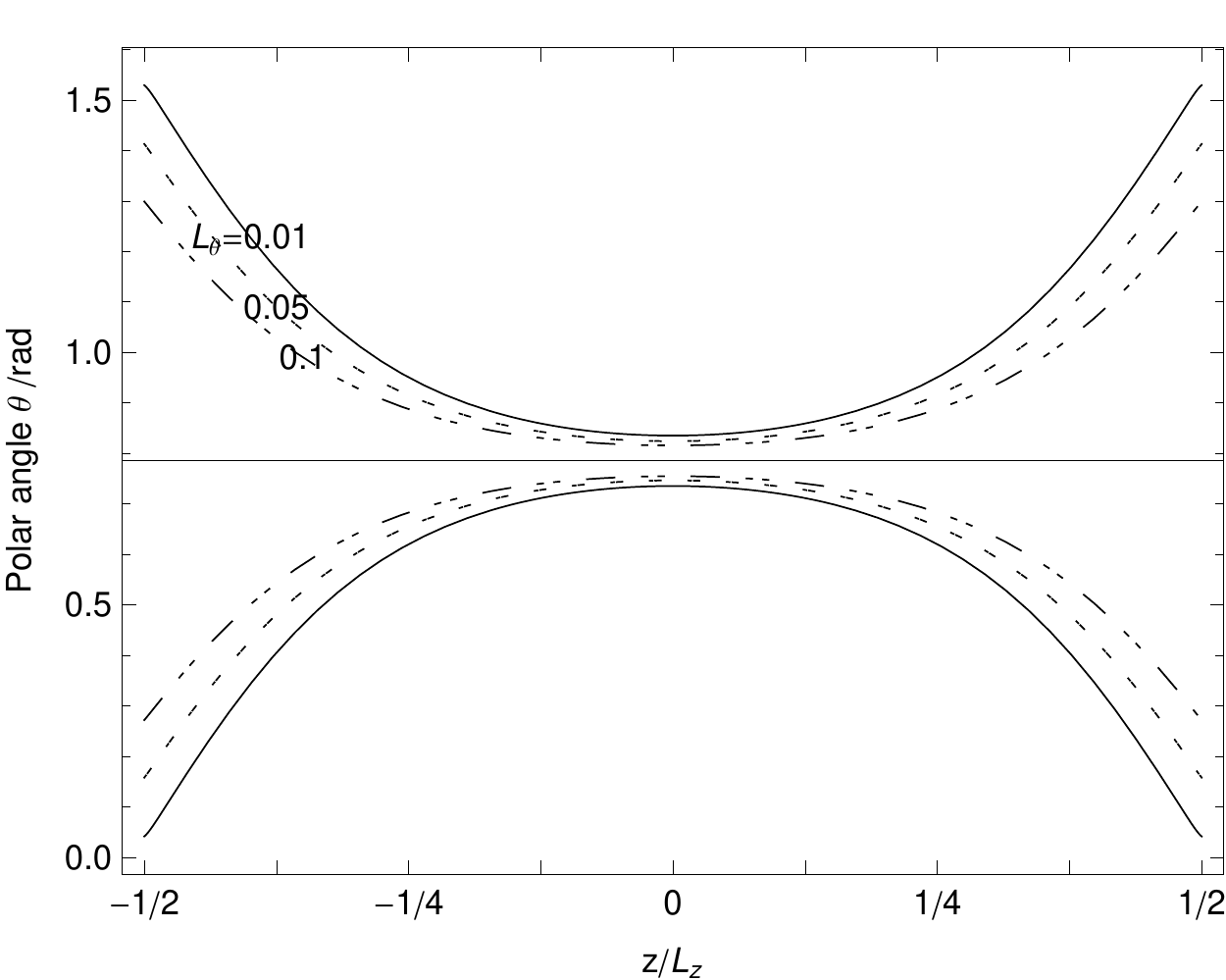}
\caption{\label{fig:zprofile}Calculated director angle profiles $\theta(z)$
in the center of the homeotropic and planar regions, for various values of $L_{\theta}$.}
\end{figure}

Continuum predictions for Director tilt profiles as a function of $z$ for
the planar and homeotropic regions are displayed in fig. \ref{fig:zprofile}
and show that a nearly uniform configuration is adopted at the cell
center, similar to the findings from our particle-based simulations. 
As the anchoring parameter $L_{\theta}$ is reduced, the tilt conditions at the substrates 
relax and the central uniform region widens

\begin{center}
\rule[0.5ex]{0.3\columnwidth}{1pt}
\par\end{center}

\section{Conclusions}
Alignment of a nematic between two substrates patterned with alternating
homeotropic and planar squares has been studied using two theoretical
approaches: A monte carlo simulation of rigid particles interacting
through the hard gaussian overlap potential and a calculation performed
with nematic continuum theory. Both techniques show a regime where
the nematic azimuthally aligns in the bulk with the edge of the squares.
In the MC simulation, the average azimuthal orientation of the molecules
is observed to flip between the two sides during the run; these states
are energetically degenerate in the continuum approach. Furthermore,
the continuum calculation reveals the existence of an anchoring transition.
If the polar anchoring is sufficiently weak and the period of the
pattern is somewhat greater than the cell thickness, the nematic instead
aligns along the diagonals. Unlike previously considered systems of
square posts and wells~\cite{Cornford2011,RefWorks:28} our calculations seem to raise the
possibility of a re-entrant behavior as a function of the period of
the pattern: with appropriate values of the polar anchoring energy,
the diagonal state becomes unstable at both short and long wavelength
patterning (fig.~\ref{fig:FreeEnergyDiff}). It is likely that the location of the critical
values of the period are only approximately correct because azimuthal
variations of the director were not included in our minimization of
the free energy. We expect that all of these transitions might be
accessible with the Monte Carlo approach for larger simulation sizes;
we are presently examining such systems. A transition between the
two regimes ought to be experimentally observable by adjusting the
ratio of the period of the pattern to the cell thickness.

Our continuum analysis also revealed the 3 result that an
exact linear form of the nematic Euler-Lagrange equations exists even
if there is 3D variation in the director and the nematic has 0
elastic constants. The form of the resulting equation lends a geometric
interpretation to elastic anisotropy as a transformation into skewed
coordinates. This result should be of utility for further study of
LC behavior in complicated geometries. 

The behavior observed in the present system is qualitatively quite
different to the case where a nematic is aligned by a striped surface.
Previous Monte Carlo sim- ulations show that for a stripe system in
a sufficiently film, there exists a regime where the nematic is divided
into domains of vertical and planar alignment that bridge the corresponding
regions of both substrates across the film. No similar behavior was
observed for the square patterned system, where instead the planar
part of the surface patterns is confined to the immediate vicinity
of the substrate. This result may be interpreted by considering the
rotation of the molecular orientation between domains, which must
occur through either a splay-like or twist-like distortion. Each of
these types of wall would have a different energy density, described
by a line tension between domains with two different values. Since
similar behavior appears from both monte carlo and continuum theory,
despite the fact that the latter entirely neglects variations in ordering,
it appears that the line tension depends primarily on elastic distortion
of the director and does not significantly depend on the scalar order
parameter. This justifies use of the continuum theory for such systems. 

In either alignment regime, whether along sides or diagonals of the
square pattern, the alignment is bistable due to the symmetry and
hence of interest for electro-optic, display and sensing applications.
In this respect these systems are quite similar to the post-aligned
bistable display~\cite{Cornford2011} or the arrays of square wells~\cite{RefWorks:28} previously
studied, which have the same symmetry, but there is an important difference:
in the present case the bistable states have no disclinations present
in the nematic configuration. Our results therefore reveal that defects
are not necessary to stabilize bistable states. Despite the apparent
simplicity of the geometry, our results illustrate the rich phase
diagrams exhibited by complex fluids in patterned geometries; further
study of these systems is presently being undertaken, to identify
optimal switching strategies between the bistable states identified
here.
\begin{acknowledgments}
This work was supported by the Engineering and Physical Research
Council, Grant No GR/S59833/01. We acknowledge useful conversations with
Steve Evans, Jim Henderson, Jon Bramble, Chris Care, Tim Spencer and Paulo Teixeira
which have been beneficial to our understanding of the systems studied here.
\end{acknowledgments}

\end{document}